\documentclass[twocolumn,showpacs,preprintnumbers,amsmath,amssymb,prl]{revtex4}

\usepackage{dcolumn}
\usepackage{bm}
\usepackage{natbib}
\usepackage{amsmath}
\usepackage{caption}

\usepackage{graphicx}

\newcommand{\kppnn}{K_{L}\rightarrow\pi^{0}\pi^{0}\nu\bar{\nu}~}
\newcommand{\kpp}{K_{L}\rightarrow\pi^{0}\pi^{0}}
\newcommand{\kppp}{K_{L}\rightarrow\pi^{0}\pi^{0}\pi^{0}}
\newcommand{\kppP}{K_{L}\rightarrow\pi^{0}\pi^{0}P}

\newcommand{\pion}{\pi^{0}}

\newcommand{\pt}{$P_{T}~$}
\newcommand{\klong}{K_L^0}
\newcommand{\mevc}{\text{MeV}/\text{c}}
\newcommand{\mevcsq}{\text{MeV}/\text{c}^2}

\newcommand*{\PUSAN}{%
$^2$Department of Physics, Pusan National University, Busan, 609-735 Republic of Korea}
\newcommand*{\SAGA}{%
$^3$Department of Physics, Saga University, Saga, 840-8502 Japan}
\newcommand*{\DUBNA}{%
$^4$Laboratory of Nuclear Problems, Joint Institute for Nuclear Research, 
Dubna, Moscow Region, 141980 Russia}
\newcommand*{\SOKENDAI}{%
$^{5}$Department of Particle and Nuclear Research, 
The Graduate University for Advanced Science (SOKENDAI), Tsukuba, Ibaraki, 305-0801 Japan}
\newcommand*{\TAIWAN}{%
$^{6}$Department of Physics, National Taiwan University, Taipei, Taiwan 10617 Republic of China}
\newcommand*{\KEK}{%
$^{7}$Institute of Particle and Nuclear Studies, 
High Energy Accelerator Research Organization (KEK), Tsukuba, Ibaraki, 305-0801 Japan}
\newcommand*{\OSAKA}{%
$^{8}$Department of Physics, Osaka University, Toyonaka, Osaka, 560-0043 Japan }
\newcommand*{\YAMAGATA}{%
$^{9}$Department of Physics, Yamagata University, Yamagata, 990-8560 Japan}
\newcommand*{\CHICAGO}{%
$^{1}$Enrico Fermi Institute, University of Chicago, Chicago, Illinois 60637, USA }
\newcommand*{\NDA}{%
$^{10}$Department of Applied Physics, National Defense Academy, Yokosuka, Kanagawa, 239-8686 Japan}
\newcommand*{\RCNP}{%
$^{11}$Research Center of Nuclear Physics, Osaka University, Ibaragi, Osaka, 567-0047 Japan}
\newcommand*{\KYOTO}{%
$^{12}$Department of Physics, Kyoto University, Kyoto, 606-8502 Japan\\ \rm (E391a collaboration) }

\begin{document}
\title{First Search for $K_{L}\rightarrow\pi^{0}\pi^{0}\nu\bar{\nu}$.}
\author{
J.~Nix$^{1}$, 
J.K.~Ahn$^2$, 
Y.~Akune$^3$, 
V.~Baranov$^4$, 
M.~Doroshenko$^{5, a}$, 
Y.~Fujioka$^3$, 
Y.B.~Hsiung$^6$, 
T.~Inagaki$^7$, 
S.~Ishibashi$^3$, 
N.~Ishihara$^7$, 
H.~Ishii$^8$, 
E.~Iwai$^8$,
T.~Iwata$^9$, 
S.~Kobayashi$^3$, 
T.K.~Komatsubara$^7$, 
A.S.~Kurilin$^4$, 
E.~Kuzmin$^4$, 
A.~Lednev$^{1, b}$, 
H.S.~Lee$^2$, 
S.Y.~Lee$^2$, 
G.Y.~Lim$^7$, 
T.~Matsumura$^{10}$, 
A.~Moisseenko$^4$, 
H.~Morii$^{12}$, 
T.~Morimoto$^7$, 
T.~Nakano$^{11}$, 
T.~Nomura$^{12}$, 
M.~Nomachi$^{8}$, 
H.~Okuno$^7$, 
K.~Omata$^7$, 
G.N.~Perdue$^{1}$, 
S.~Perov$^4$, 
S.~Podolsky$^{4, c}$, 
S.~Porokhovoy$^4$, 
K.~Sakashita$^{8,a}$, 
N.~Sasao$^{12}$, 
H.~Sato$^9$, 
T.~Sato$^7$, 
M.~Sekimoto$^7$, 
T.~Shinkawa$^{10}$, 
Y.~Sugaya$^8$, 
A.~Sugiyama$^3$, 
T.~Sumida$^{12}$, 
Y.~Tajima$^9$, 
Z.~Tsamalaidze$^4$, 
T.~Tsukamoto$^{3, *}$, 
Y.~Wah$^1$, 
H.~Watanabe$^{1, a}$, 
M.~Yamaga$^{7, d}$, 
T.~Yamanaka$^8$, 
H.Y.~Yoshida$^9$, and 
Y.~Yoshimura$^7$
\\}

\affiliation{
\CHICAGO \\
\PUSAN \\
\SAGA \\
\DUBNA \\
\SOKENDAI \\
\TAIWAN \\
\KEK \\
\OSAKA \\
\YAMAGATA \\
\NDA \\
\RCNP \\
\KYOTO \\
}
\date{\today}% It is always \today, today,

\begin{abstract}
The first search for  the rare kaon decay $\kppnn$ has been performed 
by the E391a collaboration at the KEK 12-GeV proton synchrotron. 
An upper limit of $4.7\times10^{-5}$ at the 90~\% confidence level was set 
for the branching ratio of the decay $\kppnn$ using about 10~\% of 
the data collected during the first period of data taking. First limits for the decay mode $\kppP$, where $P$ is a pseudoscalar particle, were also set.
\end{abstract}
\maketitle
%Introduction
The decay $K_{L}\rightarrow\pi^{0}\pi^{0}\nu\bar{\nu}$ is a Flavor Changing Neutral Current process \cite{Laur}  involving a $s\rightarrow d\nu\bar{\nu}$ transition. In the standard model it is a predominately CP conserving mode with the branching ratio proportional to the square of the $\rho$ parameter in the Wolfenstein parameterization of the CKM matrix. The predicted branching ratio in the standard model is $(1.4\pm0.4) \times 10^{-13}$ \cite{chiang}.   In this article, we report the first experimental limit on this mode. Additionally, we set limits on the supersymmetric mode $K_{L}\rightarrow\pi^{0}\pi^{0}P$, where $P$ is the pseudoscalar sgoldstino \cite{gorbunov}. 
 
%Sgoldstino Theory
The spontaneous breaking of any global symmetry results in a massless Nambu-Goldstone mode with the same quantum numbers as the symmetry generator. In the case of supersymmetry, the symmetry generator is fermionic resulting in a Nambu-Goldstone fermion, the goldstino. The goldstino has a superpartner, the sgoldstino, which has scalar and pseudoscalar components. If the pseudoscalar sgoldstino is light enough ($m_P < m_{K_L}-2m_{\pion}$) and the quark-sgoldstino coupling is parity conserving, then there should be the decay $\kppP$. These conditions are met in a variety of models. An upper bound on the branching ratio of this decay of  $\approx 10^{-3}$ can be derived from limits on the mass difference between $K_L$ and $K_S$ \cite{gorbunov}. Direct searches for the related charged modes $K^{\pm}\rightarrow \pi^{\pm}\pion P$ have set upper limits on those modes \cite{adlerkppp},\cite{tchikilev}. The limit on the $K^{+}$ mode is $4\times10^{-5}$ for $m_P<80 MeV/\text{c}^2$. The $K^{-}$ mode limit is $~9\times10^{-6}$ for $m_P < 200 \text{MeV}/\text{c}^2 $ . The branching ratios of the charged modes are sensitive to the magnitude of the coupling between quarks and sgoldstinos while $\kppP$ is sensitive to the real component of the coupling. Generally the $K^{\pm}\rightarrow \pi^{\pm}\pion P$ decay is suppressed by isospin conservation and therefore $\kppP$ has greater sensitivity to the coupling except when the imaginary component of the coupling dominates.

%Beam and Detector Overview
The E391a experiment at the KEK 12-GeV proton synchrotron is a dedicated experiment for the decay $K_{L}\rightarrow\pi^{0}\nu\bar{\nu}~$\cite{e391prd},\cite{thesis}. The first run of data taking took place from February to June 2004 (Run I). The analysis in this Letter corresponds to approximately 10\% of this data. The experiment uses a neutral beam extracted at $4^{\circ}$ from the primary proton line. The beam was collimated into a circular beam with a 2 mrad half cone angle \cite{beamline}. The mean $K_{L}^{0}$ momentum is 3.5 GeV/c. 

\begin{figure*}[t]
   \includegraphics*[width=0.60\textwidth,clip]{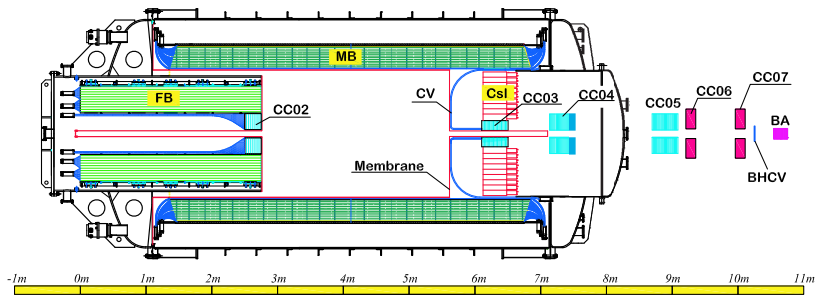}
 \caption{Cross section of the E391a detector. $\klong$'s enter from the left side.}
\label{fig:det_all}
\end{figure*}

The detector apparatus consists of a CsI crystal calorimeter and $4\pi$ hermetic photon veto system. A diagram of the detector cross section is shown in Fig. \ref{fig:det_all}. The photon vetoes are arranged cylindrically around the beamline.  The upstream end of the detector is 11 m downstream of the target, which we define as the origin of our $Z$  -coordinate. 

The calorimeter consists of 576 blocks of undoped CsI crystals \cite{csi}. The majority of crystals are $7 \times 7 \times 30 \mbox{cm}^{3} $ blocks which correspond to 16 $X_{0}$ parallel to the beam line. There are 24 $5 \times 5 \times 50 \mbox{cm}^{3}$ (=27$X_{0}$) crystals surrounding the beam hole. The energy resolution was $\sigma_{E}/E \sim 1\% / \sqrt{E} \oplus 1\%$ where $E$ is in GeV, as measured with 25 CsI crystals and a positron beam. The average position resolution of photon reconstruction is 5 mm. The face of the CsI array is located at a $Z$ position of 614.8 cm. In front of the CsI is a bank of scintillator plates (CV) for vetoing charged particles consisting of 32 overlapping 6 mm thick plastic scintillator located at 550 cm. 

There are multiple veto detectors making up the $4\pi$ photon veto system. The photon vetoes are the Front Barrel (FB), Main Barrel (MB),  collar counters (CC02-07), and Back-Anti (BA) as shown in Fig. \ref{fig:det_all}. These are all lead-scintillator sampling calorimeters, except for the BA which also has layers of quartz. The Main Barrel is the primary photon veto surrounding the fiducial decay region. It is made of lead-scintillator sandwich with an inner diameter of 2.0 m and is 13.9 $X_0$ thick. 

Most of the detectors are contained within a vacuum vessel. The fiducial decay region is kept at a pressure of $1\times10^{-5}$ Pa. This central region is separated from the rest of the detector by a 20 mg/cm$^2$ thick vacuum membrane. This membrane was not properly secured during Run I and hung into the beam line near the CV at $\approx50$ cm upstream of the CsI face. The membrane acted as a target for neutron interactions. 
%Detector Figure

%Calorimeter
%Veto Systems

The photon vetoes were calibrated using either cosmic ray or muons from the target. The CsI array was calibrated during a special run using a 5 mm Al target to produce $\pion$'s with a known position. This calibration is matched to cosmic ray muon data which is used to track changes in gain over time.

%Experimental method
The experimental signature of both $K_{L}\rightarrow\pi^{0}\pi^{0}\nu\bar{\nu}$ and $K_{L}\rightarrow\pi^{0}\pi^{0}P$ is four photons in the final state with missing mass. Their transverse momentum, $P_T$, has a higher maximum value compared to $\kpp$ or $\kppp$ with missing photons. The energy and hit position of the four photons are measured by the calorimeter. Photons are reconstructed by summing the energy of contiguous blocks with energy deposited. The position of the photon is calculated by using the distribution of visible energy in the CsI blocks. 

The possible $\pi^{0}$ decay vertices can be reconstructed assuming the decay occurred on the beam axis and the $\pion$ mass. There are multiple ways of pairing the photons to form $\pion$'s, and for each pairing we calculate a $\chi^2$ of the difference in $\pion$ $Z$ vertices. We select the pairing with the  minimum $\chi^{2}$ to determine the kaon $Z$ decay vertex. The signal modes are distinguished by their relatively high \pt and reconstructed invariant masses of the $\pion$-$\pion$ system below that of the $K_{L}$. The $\kpp$ mass peak is reconstructed with $\sigma=8$ $\text{MeV}/\text{c}^2$.

%Acceptance/Flux
The kaon flux was calculated using $K_{L}\rightarrow\pi^{0}\pi^{0}.$  It was cross-checked by $K_{L}\rightarrow\pi^{0}\pi^{0}\pi^{0}.$ The acceptance of these modes, after applying selection cuts, was calculated using a GEANT3 based Monte Carlo simulation \cite{geant3}. The simulation included an overlay of accidental events selected from data. The acceptances and fluxes for these modes are shown in Table I.

%The signal region for this mode in the \pt-mass plane is shown in Fig \ref{fig:ptmass}.

%Signal Box
The signal region was defined by the \pt of the reconstructed kaon, the invariant mass of the $\pion-\pion$ system, and the $Z$ position of the reconstructed decay vertex. The acceptable \pt was defined to be between 100 and 200 $\mbox{MeV}/\mbox{c}$. The lower bound was dictated by the presence of large amounts of $3\pion$ background at lower values of \pt. The invariant mass was required to be between 268 and 450 $\mbox{MeV}/\mbox{c}^2$. The lower bound is set by the minimal reconstructed mass with the intermediate reconstruction of two pions and the upper bound is set by the presence of $\kpp$ mass peak at the kaon's true mass of 498 $\mbox{MeV}/\mbox{c}^2$.  The acceptable decay vertices are between 300 and 500 cm. The upstream limit of 300 cm is set by the presence of beam halo neutrons interacting with CC02 at 275 cm. The downstream limit of 500 cm is set by core neutrons interacting with the vacuum membrane at $\approx 550$ cm.
%2pi0P Acceptance as function of P mass
\begin{table}[htb]
\begin{center}
\label{tbl:acceptance}
\begin{tabular}{|l|c|c|}
\hline
Mode & Acceptance & Flux \\
\hline
$\pion\pion$ &$(2.16\pm0.13)\times10^{-4}$ & $(1.54\pm0.04)\times10^{9}$\\
$\pion\pion\pion$ & $(1.39\pm0.07)\times10^{-6}$ & $(1.57\pm0.04)\times10^{9}$\\
$\pion\pion\nu\bar{\nu}$ &$(5.33\pm0.23)\times10^{-5}$ & NA\\
\hline
\end{tabular}
\end{center}
\caption{ Acceptance and flux calculations of different signal modes. Acceptance is calculated from Monte Carlo. Flux is the number of kaon decays in the fiducial region.}
\end{table}

The acceptance of the mode $\kppP$ depends on the mass of the $P$. As the mass increases there is a decline in acceptance due to the reduction in the maximum \pt of the pions. This eventually leads to the phase space of the decay to lie completely in the region of high $\kppp$ background. The single event sensitivity as a function of sgoldstino mass is shown in Fig. 2.

\begin{figure}[htbp]
   \begin{center}
    \includegraphics[width=2.7 in,clip]{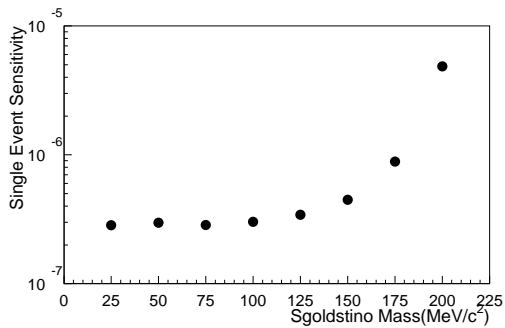}
    \end{center}
   \caption{Single event sensitivity for $\kppP$ decay as a function of sgoldstino mass. }
   \label{fig:klppPses}
\end{figure}

%Background Prediction
There are two important sources of background to the signal:  $K_{L}\rightarrow\pi^{0}\pi^{0}\pi^{0}$ with missing photons, and neutron related backgrounds. We predicted the backgrounds from data using a bifurcation technique \cite{e949bifurcation},\cite{bifurcation}. The number of observed background events can be factored into the number of events after applying a group of setup cuts and the probability of one of these events surviving the remaining cuts. If we break up the remaining cuts into two uncorrelated groupings, A and B, this probability can be factored into the probabilities of surviving each set, $P(A)$ and $P(B)$. We can rewrite this in terms of the number of events surviving the application of each set of cuts and the inverse of the other and the events which survive the application of the inverse of both cuts,
\begin{equation}
N_{\text{bkg}}=N_{A\bar{B}}N_{\bar{A}B}/N_{\bar{A}\bar{B}}.
\end{equation}

The primary source of background is $\kppp$. There are three mechanisms for this decay to produce background: two photons missing due to either geometric or detector inefficiency, one photon missing and one photon fusion (two photons reconstructed as one) in the CsI, or two photon fusions in the CsI. Background events from all three mechanisms have similar distributions in the signal space. Our cut set A consists of photon veto cuts. Cut set B is made up of cuts on the quality of the photon cluster and the particle reconstruction. 

 To check the bifurcation methodology, we applied it to regions surrounding the signal region. The Low \pt region is defined by same bounds in $Z$ and invariant mass of the $\pion$-$\pion$ system as the signal region and a \pt between 50 and 100 MeV/c. The High Mass region is defined by the same bounds in $Z$ and \pt as the signal region and a mass between 450 and 550 $\mbox{MeV}/\mbox{c}^{2}$. The High and Low $Z$ regions have the same bounds in invariant mass and \pt as the signal region and have reconstructed vertices between 500 to 550 cm and 250 to 300 cm, respectively. The predictions for these regions agree fairly well with data as shown in Table \ref{tbl:background}. 

For this technique to correctly predict the background, the cut sets A and B need to be uncorrelated. We selected cuts on this basis, but there is some correlation. An estimate of the uncertainty caused by ignoring this correlation can be made with values which are available without examining the signal region as
\begin{equation}
C_{\epsilon}= \epsilon \times N_{\bar{A}B} (1+\frac{N_{\text{pred.}}}{N_{\bar{A}B}}).
\end{equation}
Here $\epsilon$ is the difference in cut survival probability between cut A for an event passing cut B and an event passing the inverse of B. We determined $\epsilon$ using events in the Low Pt region. We estimate this as a systematic uncertainty of 0.12 background events.

One cause of cut correlation is contamination of other background sources in the signal region, primarily neutron related backgrounds. This is not measured by our estimate of $\epsilon$, because the neutron background is not present in the Low Mass, Low \pt region while it is present in the signal region. If there is a second background source with different cut survival probabilities there is a correction to the prediction. It takes the form of
\begin{equation}
N_{\text{bkg}} = \frac{ N_{A\bar{B}}    N_{\bar{A}B} }{N_{\bar{A}\bar{B}}} + \frac{N_{1}  N_{2}}{N_{\bar{A}\bar{B}}}  \Delta_{A}   \Delta_{B} \label{eqn:bkg2sol}.
\end{equation}
 Here, $N_{1}$ and $N_{2}$ are the number of events of each background type before cuts A and B are applied, and $\Delta_{A}$ and $\Delta_{B}$ are the difference in the veto probabilities between the background types for the two cuts. It is important to note that the correction term does not directly correspond to the background contribution from the secondary sources. 

The largest source of neutrons is the interaction of beam core neutrons with the vacuum membrane in front of the CV. A second source is the interaction of halo neutrons with CC02. Both of the sources produce reconstructed events which are localized to their point of interaction with high $P_{T}.$ Both of these regions are outside the fiducial decay region. To estimate the impact of these events on the the background prediction, it was necessary to determine the number of core neutron background events in the signal region under the setup cuts. 

The number of neutron events in the region with $Z$  greater than 500 cm and $P_{T}$ greater than 0.1 GeV/c is too small to fit when all cuts are applied. We therefore fit the data using a loose set of cuts. We remove the BA cut, cuts on the distribution of photon energy and timing, and a veto on additional energy in the CsI. Additionally, we apply the inverses of cuts A and B to ensure we are not observing events in the signal region while being able to look at events in the $Z$ fiducial region. With these sets of cuts the core neutron peak in the high \pt-high $Z$ region is clearly visible and can be fit with a Gaussian as shown in Fig. 3. In the high \pt-fiducial $Z$ region, there is a predominately $\kppp$ background. The Gaussian component of the core neutron peak produces a negligible contribution to the background in the fiducial region. The density of this distribution is calculated by subtracting off the $\kppp$ contribution found by Monte Carlo simulation and fitting the remainder by a Gaussian plus a straight line. Integrating this function over the fiducial region gives $248\pm 4_{\text{stat.}}\pm124_{\text{syst.}}$ core neutron background events with the loose cuts applied. The application of the rest of the setup cuts reduces this by a factor of $\approx 115$. A predicted total core neutron background in the signal region is $2.16\pm0.03_{\text{stat.}}\pm1.05_{\text{syst.}}$ under the setup cuts, before the application of cuts A and B. This corresponds to $N_{2}$ in Equation \ref{eqn:bkg2sol}. 

The cut survival probabilities do differ significantly with $\Delta_{A}=(26.4\pm1.3)\%$ and $\Delta_{B}=(8.6\pm0.5)\%$ which are derived from Monte Carlo studies. $N_{1}$ is estimated by assuming the background before the setup cuts are applied is dominated by $\kppp$ events and subtracting off the core neutron contribution. We estimate $N_{1}=102\pm 10$. This results in an estimation of the uncertainty to the background prediction of $0.06$ events. 
\begin{figure}[htbp]
   \begin{center}
    \includegraphics[width=2.7 in,clip]{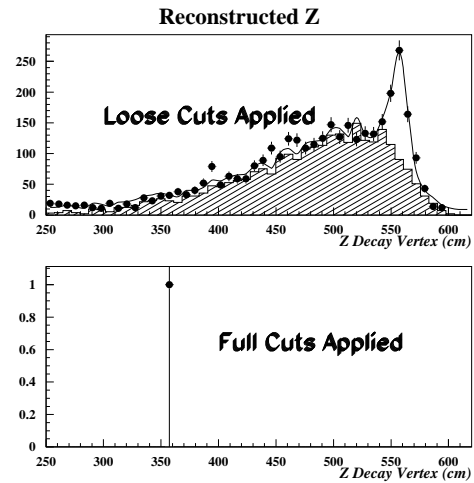}
    \end{center}
   \caption{$Z$ distribution of reconstructed events with $P_T>100$ MeV/c. Points are data, shaded region represents a scaled $\kppp$ Monte Carlo sample, and the line is the fitted curve. The top plot shows the distribution with the loose set of cuts from which we derived our neutron background prediction, and the bottom plot shows the same distribution with all cuts applied. }
   \label{fig:coreneutron}
\end{figure}

\begin{table}[htb]
\caption{Prediction of background events in different regions. Statistical errors are shown.}
\begin{center}
\label{tbl:background}
\begin{tabular}{|c|c|c|c|c|c|}
\hline
 & & & & & \\
Region & $N_{\bar{A}\bar{B}}$ &$N_{A\bar{B}}$ & $N_{\bar{A} B} $ & Prediction & Data \\
\hline
Low \pt & $393$ & $72$ & $115$ & $21.1\pm3.3$ & 13  \\
\hline
High Mass & $46$ & $9$ & $4$ & $0.78\pm0.48$ & 1 \\
\hline
Low $Z$  & $5$ & 0 & 0 & 0 & 0 \\
\hline
High $Z$  & $0$ & 0 & $6$ & 0 & 0 \\
\hline
Signal & $84$ & $18$ & $2$ & $0.43\pm0.32$ & 1 \\
\hline
\end{tabular}
\end{center}
\end{table}

\begin{figure}[htbp]
  \begin{center}
    \includegraphics[width=2.7 in,angle=90,clip]{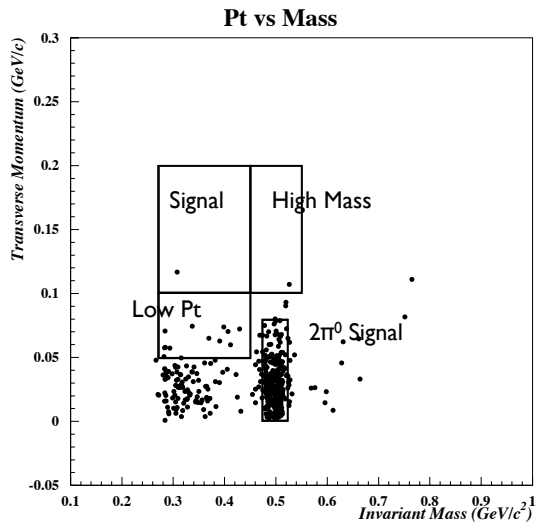}
   \end{center}
   \caption{\pt plotted versus invariant mass with all selection cuts applied. The rectangular regions correspond to the regions in Table \ref{tbl:background}. The $2\pi^0$ signal region was used to calculate the kaon flux in Table I. }
   \label{fig:ptmass}
 \end{figure}
 
The bifurcation background predictions gives us a total background prediction of $0.43\pm0.32_{\text{stat.}}\pm0.13_{\text{sys.}}$. In the signal region we observe a single event shown in Fig. \ref{fig:ptmass}. The event has $M_{\pion-\pion}=315 \pm 8 ~\mevcsq$, $P_{T}=111 \pm5~ \mevc$, and $Z=357\pm 11 ~\text{cm}$. This is consistent with the background prediction. 
 
The uncertainty in the branching ratio of $\kpp$ contributes a systematic uncertainty of 0.5\% to our single event sensitivity. Uncertainty in the calibration of the photon vetoes contributes a systematic uncertainty of $3.7\%$.  Discrepancies between Monte Carlo and data acceptance loss in the cuts, which were simulated by Monte Carlo, give a systematic error of $3.4\%$ in acceptance. The total systematic acceptance error is $5.0\%$. Combining the branching ratio and acceptance errors produces a $7.1\%$ systematic error in the single event sensitivity.

Our single event sensitivity for $\kppnn$is $(1.20\pm0.06_{\text{stat.}}\pm0.09_{\text{sys.}})\times10^{-5}$. With 1 observed event, which is consistent with the background prediction, we set a limit at the 90\% confidence level for the branching ratio at $4.7\times10^{-5}$ using Poisson statistics. This is the first limit set on this decay mode. The single event sensitivities for $\kppP$ with different sgoldstino masses are shown in Fig. \ref{fig:klppPses}. For $m_P< 100 \text{MeV}/\mbox{c}^2$, we can set a limit on $\kppP$ of $1.2\times10^{-6}$ at the 90\% confidence level.
 
This analysis uses $~10\%$ of the data from the first E391a run. It appears that with this data set the analysis is background limited. We expect the second and third runs of E391a to have significantly reduced neutron backgrounds due to correcting the vacuum membrane issue and an insertion of a Be absorber into the beam. The second and third runs have more than 10 times the data as this analysis. However, further study is needed to determine the proper cut selection and what acceptance can be achieved in this improved neutron environment. 

%Acknowledgement
We are grateful to the operating crew of the KEK 12-GeV proton synchrotron for their successful beam operation during the experiment. This work has been partly supported by a Grant-in-Aid from the MEXT and JSPS in Japan, a grant from NSC in Taiwan, a grant from KRF in Korea,  the U.S. Department of Energy, and the Grainger Foundation.

\noindent  
$^*$Deceased \\
$^a$Present address: KEK, Tsukuba, Ibaraki, 305-0801 Japan. \\
$^b$Also Institute for High Energy Physics, Protvino, Moscow region, 142281 Russia. \\
$^c$Also Scarina Gomel' State University, Gomel', BY-246699, Belarus. \\
$^d$Present address:  Osaka University, Toyonaka, Osaka, 560-0043 Japan.

\bibliographystyle{plain}

\end{document}